# Gallium substitution in $PuCoGa_5$

E. Colineau[a*], P. Boulet[b], J.-C. Griveau[a], R. Eloirdi[a], J. Rebizant[a], F. Wastin[a,+], A.B. Shick[c] and R. Caciuffo[a]

*[a]European Commission, Joint Research Centre (JRC), Institute for Transuranium Elements (ITU), Postfach 2340, 76125 Karlsruhe, Germany*

*[b]Institut Jean Lamour, UMR 7198 CNRS-Université de lorraine, Parc de Saurupt, 54011 Nancy cedex, France*

*[c]Institute of Physics, ASCR, Na Slovance 2, CZ-18221 Prague, Czech Republic*

*[+]Present address: European Commission, Joint Research Centre (JRC), Institute for Energy and Transport (IET), P.O. Box 2, 1755 ZG Petten, The Netherlands*

Abstract

The substitution of gallium by aluminum, germanium, tin and indium in $PuCoGa_5$, the actinide-based superconductor with the highest critical temperature, has been investigated. Only systems with 20% substitution by Al and Ge (i.e. $PuCoGa_4Al$ and $PuCoGa_4Ge$) have been successfully synthesized by annealing arc-melted samples. X-ray powder diffraction refinements indicate an enhanced c/a ratio in these two compounds but the magnetic susceptibility measurements reveal a large reduction of the critical temperature $T_c$ compared to $PuCoGa_5$. DFT+ED calculations indicate significant changes in the Fermi surface, probably related to the decrease of $T_c$, in these compounds. In addition, the isotopic effect of the atomic mass of gallium, using $^{69}Ga$ and $^{71}Ga$ isotopes, was also investigated without observable influence on the critical temperature of $PuCoGa_5$.

*Corresponding author : E. Colineau

Email : eric.colineau@ec.europa.eu

Tel. : +49 (0)7247 951 442

Fax : +49 (0)7247 951 99 442

**I- Introduction**

The discovery of superconductivity in $PuCoGa_5$ with a critical temperature one order of magnitude larger than those observed in uranium heavy fermion superconductors [1] has led to intensive experimental and theoretical works on this unique material. Evidences for unconventional superconductivity with d-wave symmetry have been obtained [2-5] but the pairing mechanism still remains unclear [6]. It seems that $PuCoGa_5$ is not close to a magnetic instability [7- 9].

The lattice parameters and electronic structure of $PuCoGa_5$ have been tuned by applied pressure and chemical substitution: the critical temperature has been enhanced from $T_c$ = 18.5 K up to a maximum of $T_c$ = 21.5 K under applied pressure [10]. On the contrary, no enhancement of temperature could be achieved with any chemical substitution, which instead induces a rapid decrease, even with small amounts of doping. Plutonium has been substituted by uranium, neptunium [7] and americium [11, 12], cobalt has been substituted by rhodium, iron, nickel [7], but gallium substitution had not been reported yet, except the unique case of fully substituted $PuCoIn_5$ [9]. In this paper, we report on gallium substitution by aluminum, germanium, and tin, where only 20% Al and Ge substitutions were successful. We also show the absence of isotopic effect of gallium in $PuCoGa_5$.

**II- Experimental**

Polycrystalline ingot of all the compounds discussed hereafter were obtained by arc melting stoichiometric amounts of the constituent elements under an atmosphere of high purity argon on a water-cooled copper hearth, using a Zr getter. Starting materials were used in the form of 3N8 shot of cobalt, 3N7 shot of metalloid element i.e. Ga, Ge, Al, and Sn as supplied by A.D. Mackay Inc., and 3N6 plutonium metal. Homogeneity of the sample was ensured by turning over and re-melting the button several times. Weight losses were below 0.5%.

In agreement with the previous studies on the $PuCoGa_5$ system [1, 7] the arc melted samples were annealed at 750°C for a week to obtain the pure phase. The phase purity of the sample was checked by X-ray powder diffraction data (Cu Kα radiation) collected on a Bragg-Brentano Siemens D500 diffractometer using a 2θ step size of 0.02 degrees. The diffraction patterns were analyzed by a Rietveld-type profile refinement method using the Fullprof program [13]. Magnetization measurements have been performed on a QD-SQUID magnetometer at temperatures down to 2 K and with applied magnetic fields up to 7 T. The masses of the $PuCoGa_4Al$, $PuCoGa_4Ge$, $PuCo^{69}Ga_5$ and $PuCo^{71}Ga_5$ polycrystalline samples were 135, 162, 59 and 33 mg, respectively. All samples have been measured immediately after synthesis, to avoid ageing effects due to the α-decay of the $^{239}$Pu isotope [14, 15].

## III- Results

### 1-Structural Chemistry:

Since the discovery of superconductivity in $PuCoGa_5$, many attempts to obtain isostructural phases by changing one of the constituting elements have been performed to tune the superconducting properties. These attempts have been successfully performed by replacing plutonium with other actinides elements, such as U, Np, or Am. Cobalt can also be replaced by other transition metals such as Fe, Ni, Rh or Ir. Up to now, and except for $PuCoIn_5$ which was obtained by self-flux method, no study on the substitution of gallium atoms have been reported.

Figure 1 shows the X-ray powder diffraction pattern obtained for a sample of $PuCoGa_5$ prepared with the $^{69}$Ga isotope, after annealing at 750°C of the arc melted button. The obtained unit cell parameters are shown in table 1 and are very close to those reported for $PuCoGa_5$.

Whilst Ga isotope substitution is easy to obtain, attempts to prepare $PuCoX_5$ (X=Al, Ge, In and Sn) using the arc melting procedure developed for the synthesis of $PuCoGa_5$ failed. In all cases, the results show crystallization of the pseudo-binary $PuCo_xX_{3-x}$ as a major phase adopting in the four cases the cubic $AuCu_3$ type, with structure lattice parameters slightly larger than those reported for the pure binary, i.e. 4.275 Å, 4.245 Å and 4.639 Å respectively for $PuCo_xAl_{3-x}$, $PuCo_xGe_{3-x}$ and $PuCo_xSn_{3-x}$. Other diffraction peaks with lower intensity reveal the presence of β–Sn and $CoSn_3$ in the case of the Sn sample. It should be mentioned here that the $AuCu_3$ structure type is highly favourable in the case of Al, Ge and Sn based systems since $PuAl_3$ [16], $PuGe_3$ [17], $PuIn_3$ [18], and $PuSn_3$ [19] are known to crystallize with the $AuCu_3$ type contrary to $PuGa_3$ [20], which adopts the two polymorph structures $Mg_3In$ (R-3m) and $Mg_3Cd$ ($P6_3/mmc$).

However, successful preparations using the arc-melting procedure have been obtained when replacing 20% of Ga atoms by aluminum or germanium as shown by the Rietveld refinement in Figure 2. The corresponding lattice parameters are listed in table 1. The crystal structure is displayed in Figure 3, showing in particular the atomic position of Ga with the Wyckoff position 4i and Ge in 1c. These results revealed that the phase $PuCoGa_4Al$ and $PuCoGa_4Ge$ exist and could be prepared by arc melting. The crystal structure adopted is the $HoCoGa_5$ type as for $PuCoGa_5$. These results and the possible reasons to explain why Ga substitution is obtained only up to 20 % are discussed in section 4.

**2- Physical properties:**

As shown in Figure 4, the volume magnetic susceptibility of $PuCoGa_4Al$ undergoes a sharp collapse at $T_c \approx 12$ K ($T_{onset} = 12.2$ K) from $\chi \approx 2.6\times10^{-4}$ down to $\chi \approx -1$ (SI), clearly indicating the onset of bulk superconductivity in this sample. A small anomaly in the transition is observed around 11.8 K and could be related to a slight inhomogeneity of the sample.

The magnetization as function of magnetic field (Figure 5) is irreversible and evidences flux pinning, similarly to $PuCoGa_5$ but with lower critical parameters: $\mu_0H_{c1}$ is estimated to about 0.01 T and $\mu_0H_{c2}$ is probably close to 7 T, i.e. a value comparable to that of $Pu_{0.9}U_{0.1}CoGa_5$ (8.7 T [7]) and one order of magnitude smaller than in $PuCoGa_5$ (~ 100 T [6]).

$PuCoGa_4Ge$ also shows a sudden collapse of its magnetic susceptibility (Fig. 6) but much less sharp and pronounced than its $PuCoGa_4Al$ counterpart: the onset of the decrease is observed at $T_{onset} = 7.5$ K, but it only decreases from $\chi \approx 4\times10^{-4}$ down to $\chi \approx -0.2\times10^{-3}$. If we interpret this collapse as a superconducting transition, this means that only 0.02% of the sample volume would be superconducting at 3 K (the lowest achieved temperature on this sample). The superconducting fraction might increase at lower temperature, however, the transition is unusually broad (from 7.5 K to less than 3 K), with a broad anomaly around 5 K, suggesting that the sample is not homogenous. The field dependence of the magnetization (Figure 7) is consistent with a very small fraction of superconducting material (with $\mu_0H_{c1} < 0.01$ T) in a predominantly paramagnetic sample.

In summary, we observe clear superconductivity in the whole sample volume of $PuCoGa_4Al$ below $T_c \approx 12$ K and possible superconductivity below $T_c \approx 7$ K in a small fraction of the $PuCoGa_4Ge$ sample. In both samples, no anomaly can be detected around 18-19 K, which means that these samples do not contain any trace of the $PuCoGa_5$ phase whose superconducting signal ($T_c \approx 18.6$ K) would otherwise clearly emerge.

Finally, we have measured the magnetic susceptibility of $PuCoGa_5$ made with two different isotopes of gallium, $^{69}Ga$ and $^{71}Ga$. In the BCS theory the critical temperature is proportional to the inverse of the square root of the molar mass. The molar mass difference between $PuCo^{69}Ga_5$ ($M_1 = 643.13$ g/mol) and $PuCo^{71}Ga_5$ ($M_2 = 653.13$ g/mol) is 10 g/mol, i.e. we expect $T_{c1}/T_{c2} = (M_2/M_1)^{1/2} = 1.008$. We observe that both onset temperatures (and temperatures taken at half transition) are identical, i.e. $T_{c1} = T_{c2}$ ($\pm$0.1K), but the experimental error is only slightly inferior to the expected 0.8% (0.14 K) isotopic effect, therefore this result should still be considered with care.

## 3- Electronic Structure Calculations

To examine theoretically the electronic structure of $PuCoGa_5$, $PuCoGa_4Al$, and $PuCoGa_4Ge$ we performed DFT+"Exact Diagonalization" (ED) calculations, as explained in details in [5, 25, 26]. In this approach, the band structure obtained by the relativistic version of the full-potential linearized augmented plane wave method (FP-LAPW) [27] is combined with the many-body solution of a discretised single-impurity Anderson model in order to account for the full structure of the 5f-orbital atomic multiplets and their hybridization with the conduction bands.

The Coulomb interaction in the Pu-atom 5*f* shell is parameterized by Slater integrals $F_0$ = 4.00 eV, $F_2$ = 7.76 eV, $F_4$ = 5.05 eV and $F_6$ = 3.70 eV for the Pu 5*f* shell, as given in [28]. All calculations are performed using the experimental lattice parameters and the internal atomic positions listed in Table 1.

Calculated occupancies of the f-shell ($n_f$) ground state in $PuCoGa_5$, $PuCoGa_4Al$, and $PuCoGa_4Ge$ are shown in Table 2, and are close to 5.3. This corresponds to the intermediate valence of the Pu atom in these compounds. The probabilities of finding the Pu atom in $f^4$, $f^5$, $f^6$, $f^7$ integer valence configurations are listed in Table 2, and show that the $f^5$ and $f^6$ configurations contribute the most into the ground state. No significant $f^4$ content is calculated, contrary to the suggestion of Booth *et al.* [29], and in agreement with recent DMFT calculations of $PuCoGa_5$ [30]. The incorrect conclusions made in [29] are probably due to the use of obsolete quantum-chemical techniques which completely neglect the hybridization of the localized f-state with the bath of s, p, and d conduction electrons.

This hybridization, accounted for in DFT+DMFT [30] as well as in the present DFT+ED calculations, plays a crucial role in the non-magnetic behaviour of these materials. The many-body ground state of the cluster formed by the 5f shell and the bath is given by a non-magnetic singlet with all angular moments of the 5f-bath cluster equal to zero.

In Figures 8-10, we show the mean-field band structure and Fermi Surface (FS) for $PuCoGa_5$, $PuCoGa_4Al$, and $PuCoGa_4Ge$. Note that the FS plots are centred around the Γ-point, which is hidden behind the FS sheets. In the case of isoelectronic Al-substitution, the band structure and FS looks rather similar to the parent material. There is a pronounced decrease in the “hybridization gap” between FS1 and FS2 bands along the Z-A direction.

As the mechanism of superconductivity in the Pu "1:1:5" family is still under debate, we can only speculate that the decrease of the critical temperature $T_c$ is connected to the change in FS-2 sheet for $PuCoGa_4Al$ (cf., Fig. 8 and Fig. 9).

In the case of $PuCoGa_4Ge$, where Ge adds an additional p-electron, the change in FS is more pronounced. The FS-1 sheet of the holes around the Γ-point is not seen in Fig. 10, as it is covered by FS-2. The FS-2 sheet now consists of 2 pieces, as well as the FS-3 sheet. The FS-4 sheet is similar to the cases of $PuCoGa_5$ and $PuCoGa_4Al$.

**4- Discussion**

The crystal structure adopted by the $AnTGa_5$ compounds is a tetragonal structure with space group P4/mmm where the actinide and metal atoms each occupy only one crystallographic site, i.e. the Wyckoff position 1a (0 0 0) and 1b (0 0 ½), respectively. Both are special positions fixed by symmetry. On the contrary, the metalloid element occupies two distinct crystallographic sites: the first one (Ga1) is a fixed position 1c (½ ½ 0) and the second (Ga2) is the less symmetric 4i Wyckoff position (0 ½ z), with one variable along z. The 1c site is surrounded by 12 atoms and corresponds to a cubic site formed by 8 Ga2 at 2.98 Å, for $PuCoGa_5$, capped by 4 Pu atoms forming a square on the basal plane at 2.99 Å. The second metalloid site is obviously of lower symmetry with 11 surrounding atoms and consisting of a pentagonal plane formed by 2 plutonium, 2 metals and 1 Ga2, bi-capped by 2 Gallium triangles. The interatomic distances span from 2.48 Å to 2.98 Å, i.e. smaller than those observed for the first Ga site.

In the case of plutonium or cobalt substitution by another element, only one crystallographic site is involved and it does not affect much the unit cell [7]. In the case of metalloid substitution, two different crystallographic sites are involved, with different environments. As explained above, the site 1c, corresponding to 20% of Ga, is larger than the other one, which could explain why it appears easier to obtain the 20% Ga substitution as a first step. Further substitution (larger than 20%) of Ga by another atom probably requires more energy, not provided by annealing at 750°C. However, the flux method is known to decrease the energy barrier which might explain why complete substitution of Ga by In can be achieved by indium self-flux.

The effects on superconductivity of gallium substitution in $PuCoGa_5$ appear to be strongly dependent on the substitute: aluminum decreases the critical temperature by 6.6 K, whereas germanium seems to decrease it much more dramatically (more than 11 K, at least). This conclusion is to be compared with cobalt substitution [7] where isoelectronic substitution was observed to be much less destructive for superconductivity. In the case of gallium substitution, the same tendency is observed: aluminum is isoelectronic to gallium, whereas germanium has one more p electron. Our calculations show that the Fermi surface follows the same trend – more dramatic changes with non-isoelectronic substitution - and is probably connected to the change of the critical temperature.

If we refer to the cerium analogues [2] or the actinide or cobalt substitutions [7], where the critical temperature is observed to increase with the lattice parameters c/a ratio, we would expect an increase of $T_c$ from $PuCoGa_5$ (c/a = 1.604) to $PuCoGa_4Al$ and $PuCoGa_4Ge$ which exhibit c/a = 1.609 and c/a = 1.638, respectively. Instead, we clearly observe the opposite trend, like in the case of the fully substituted $PuCoIn_5$ ($T_c$ = 2.5 K and c/a = 1.626) [9]. Table 1 lists the lattice parameters and critical temperature of $PuCoGa_5$ and Pu-, Co- or Ga-substituted analogues.

Even in the "favourable" cases of isoelectronic substitution or enhanced c/a ratio, the critical temperature of $PuCoGa_5$ cannot be enhanced or even preserved. The exact chemical composition of this ternary compound seems to be an optimum for superconductivity in this system - the critical temperature can be increased only by applied pressure (figure 11).

## IV- Conclusion

The substitution of gallium in $PuCoGa_5$ by other elements from groups IIA and IVA has been investigated. Stoichiometric $PuCoAl_5$, $PuCoGe_5$, $PuCoSn_5$ and $PuCoIn_5$ could not be obtained by arc melting and only synthesis with aluminium and germanium, in the proportion of 1 to 5 atoms of gallium, corresponding to one of the two crystallographic sites occupied by gallium in the unit cell, could be successfully performed.

$PuCoGa_4Al$ shows bulk superconductivity below 12 K whereas $PuCoGa_4Ge$ shows some hints of superconductivity below 7 K. This confirms the trend previously observed (by actinide and cobalt substitution) that isoelectronic substitution is less destructive for superconductivity. Calculations show the same trend for the Fermi surface, which is clearly less affected by isoelectronic substitution.

However, the trend to enhanced superconductivity with larger c/a ratio is here contradicted since both samples have larger c/a ratio but smaller critical temperatures than $PuCoGa_5$. Finally, we also showed that the isotopic mass of gallium has no observable influence on the critical temperature of $PuCoGa_5$.

**Acknowledgements:**

The authors acknowledge F. Kinnart and D. Bouëxière for their technical support. The high-purity plutonium metal required for the fabrication of the compounds was made available through a loan agreement between Lawrence Livermore National Laboratory and ITU, in the frame of a collaboration involving LLNL, Los Alamos National Laboratory, and the U. S. Department of Energy. P. Boulet acknowledges the collaboration with JRC-ITU in the frame of EARL. A.B.S. acknowledges support from the Czech Science Foundation (GACR) Grant No. 15-07172S, and access to computing and storage facilities of the National Grid Infrastructure MetaCentrum, provided under the program LM2010005.

**Table 1**: Lattice parameters (a, c and z atomic coordinate of Ga (4i)) determined by Rietveld refinement of the corresponding XRD powder pattern and critical temperature ($T_c$) of $PuCoGa_5$ and substituted compounds.

| Substitution | Compound | a (Å) | c (Å) | $z_{Ga2}$ | c/a | $T_c$ (K) | Reference |
|---|---|---|---|---|---|---|---|
| - | $PuCoGa_5$ (P ~ 10-15 GPa) | 3.968(1) | 6.406(1) | | 1.619 | 21.5 | 2, 10, 21 |
| | $PuCoGa_5$ | 4.2354(1) | 6.7953(3) | 0.3091(4) | 1.604 | 18.5 | 1,23 |
| Pu | $Pu_{0.9}U_{0.1}CoGa_5$ | 4.2281(1) | 6.7784(1) | 0.3088(3) | 1.603 | 8.4 | 7 |
| | $Pu_{0.9}Np_{0.1}CoGa_5$ | 4.2296(2) | 6.7769(4) | 0.3070(6) | 1.602 | 7.2 | 7 |
| | $AmCoGa_5$ | 4.2328(1) | 6.8241(2) | 0.3106(5) | 1.612 | 1.9 | 11 |
| | $Pu_{0.88}Am_{0.12}CoGa_5$ | 4.2297(2) | 6.7834(1) | 0.3081(6) | 1.604 | <6.5K[a] | 12 |
| | $Pu_{0.8}Am_{0.2}CoGa_5$ | 4.2287(1) | 6.7862(1) | 0.3079(3) | 1.604 | <2K[a] | 12 |
| | $NpCoGa_5$ | 4.2377(1) | 6.7871(2) | 0.3103(4) | 1.601 | <0.4K[a] | 22 |
| | $UCoGa_5$ | 4.2446(1) | 6.7410(2) | 0.3075(3) | 1.588 | <1.8K[a] | 7,22 |
| | $PuCo_{0.9}Ni_{0.1}Ga_5$ | 4.2294(3) | 6.7792(6) | 0.3089(5) | 1.602 | 16.6 | 7 |
| Co | $PuCo_{0.5}Rh_{0.5}Ga_5$ | 4.2610(2) | 6.8185(4) | 0.3069(8) | 1.600 | 15.5 | 7 |
| | $PuCo_{0.9}Fe_{0.1}Ga_5$ | 4.2307(1) | 6.7805(4) | 0.3093(8) | 1.602 | 13.5 | 7 |
| | $PuCo_{0.1}Rh_{0.9}Ga_5$ | 4.2931(1) | 6.8495(2) | 0.3034(5) | 1.595 | 10.2 | 7 |
| | $PuCo_{0.8}Fe_{0.2}Ga_5$ | 4.2329(2) | 6.7809(4) | 0.309(1) | 1.601 | 10.0 | 7 |
| | $PuRhGa_5$ | 4.3012(1) | 6.8570(2) | 0.3064(3) | 1.594 | 8.9 | 23 |
| | $PuIrGa_5$ | 4.3243(1) | 6.8178(2) | 0.3038(4) | 1.576 | <2K[a] | 24 |
| | $PuFeGa_5$ | 4.2685(7) | 6.7521(8) | 0.259(3) | 1.582 | <2K[a] | 7 |
| | $PuNiGa_5$ | 4.2452(1) | 6.7963(2) | 0.3070(3) | 1.600 | <2K[a] | 7 |
| Ga | $PuCoGa_4Al$ | 4.2203(1) | 6.7915(4) | 0.322(1) | 1.609 | 12.2 | This work |
| | $PuCoGa_4Ge$ | 4.1905(1) | 6.8629(2) | 0.3117(3) | 1.638 | ~7 | This work |
| | $PuCo^{69}Ga_5$ | 4.2369(1) | 6.7955(3) | 0.3095(4) | 1.603 | | This work |
| | $PuCo^{71}Ga_5$ | 4.2366(1) | 6.7952(3) | 0.3084(4) | 1.603 | | This work |

[a] No superconductivity was observed down to the temperature indicated (lowest experimental temperature achieved).

**Table 2:** Occupancy of the f-manifold ($n_f$) ground state, and $f^4$, $f^5$, $f^6$, $f^7$ –components content in the ground state of Pu atom in $PuCoGa_5$, $PuCoGa_4Al$, and $PuCoGa_4Ge$.

| | $n_f$ | $f^4$ | $f^5$ | $f^6$ | $f^7$ |
|---|---|---|---|---|---|
| **$PuCoGa_5$** | 5.29 | 0.03 | 0.65 | 0.31 | 0.01 |
| **$PuCoGa_4Al$** | 5.29 | 0.03 | 0.66 | 0.30 | 0.01 |
| **$PuCoGa_4Ge$** | 5.31 | 0.04 | 0.63 | 0.33 | 0.01 |

Figure captions:

**Figure 1**: (Color online) X-ray powder diffraction refinement of the $PuCo^{69}Ga_5$ sample. The dots represent the observed data points; the solid lines reveal the calculated profile and the difference between observed and calculated profiles. The ticks correspond to the $2\theta_{hkl}$ Bragg positions of the $PuCo^{69}Ga_5$ phase.

**Figure 2**: (Color online) X-ray powder diffraction refinement of the $PuCoGa_4Ge$ sample. The dots represent the observed data points; the solid lines reveal the calculated profile and the difference between observed and calculated profiles. The ticks correspond to the $2\theta_{hkl}$ Bragg positions of the $PuCoGa_4Ge$ phase.

**Figure 3**: (Color online) View of the $PuCoGa_4Ge$ crystal structure, with the Germanium atoms (Wyckoff position 1c) in the basal plane of the tetragonal unit cell, i.e. in the same plane than the Pu atoms. The Ga atoms (Wyckoff position 4i) occupy two layer planes above and below the plutonium layer.

**Figure 4**: (Color online) Magnetic susceptibility of $PuCoGa_4Al$ as a function of temperature.

**Figure 5**: (Color online) Magnetization of $PuCoGa_4Al$ as a function of magnetic field. Inset: zoom at low field.

**Figure 6**: (Color online) Magnetic susceptibility of $PuCoGa_4Ge$ as a function of temperature.

**Figure 7**: (Color online) Magnetization of $PuCoGa_4Ge$ as a function of magnetic field. Inset: zoom at low field.

**Figure 8:** (Color online) $PuCoGa_5$ band structure where the circles indicate the weighted f-character of the electronic states, and Fermi Surface.

**Figure 9:** (Color online) $PuCoGa_4Al$ band structure and Fermi Surface**.**

**Figure 10:** (Color online) $PuCoGa_4Ge$ band structure and Fermi Surface.

**Figure 11:** (Color online) Critical temperature $T_c$ and lattice parameters ratio c/a for different compounds related to $PuCoGa_5$: Co-substituted (green area, left), Pu-substituted

(yellow area, center) and Ga-substituted (blue area, right). Circles indicate superconducting compounds, whereas squares denote compounds where no superconductivity was detected down to the lowest experimental temperature (typically 2K).

Figures:

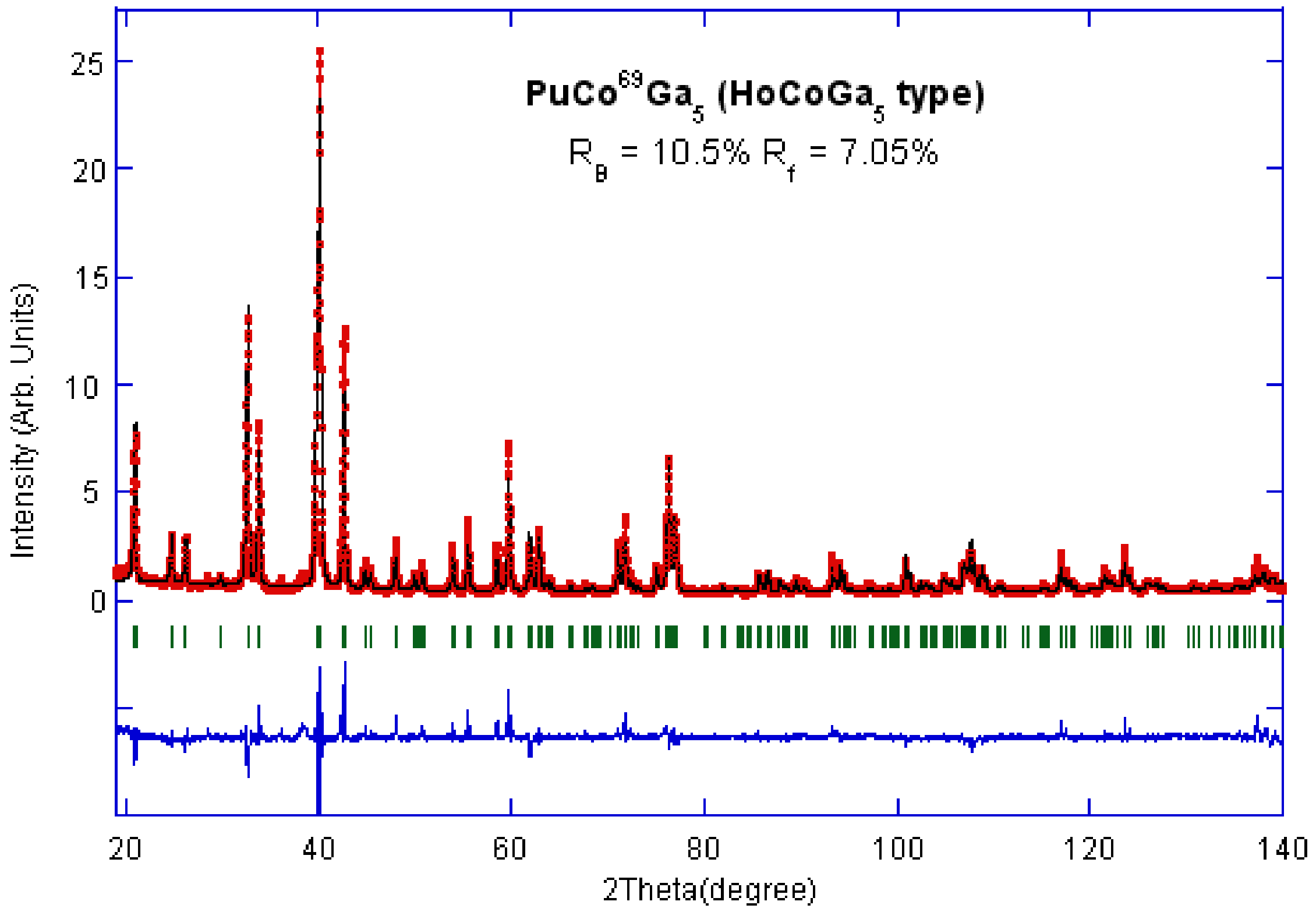

Figure 1

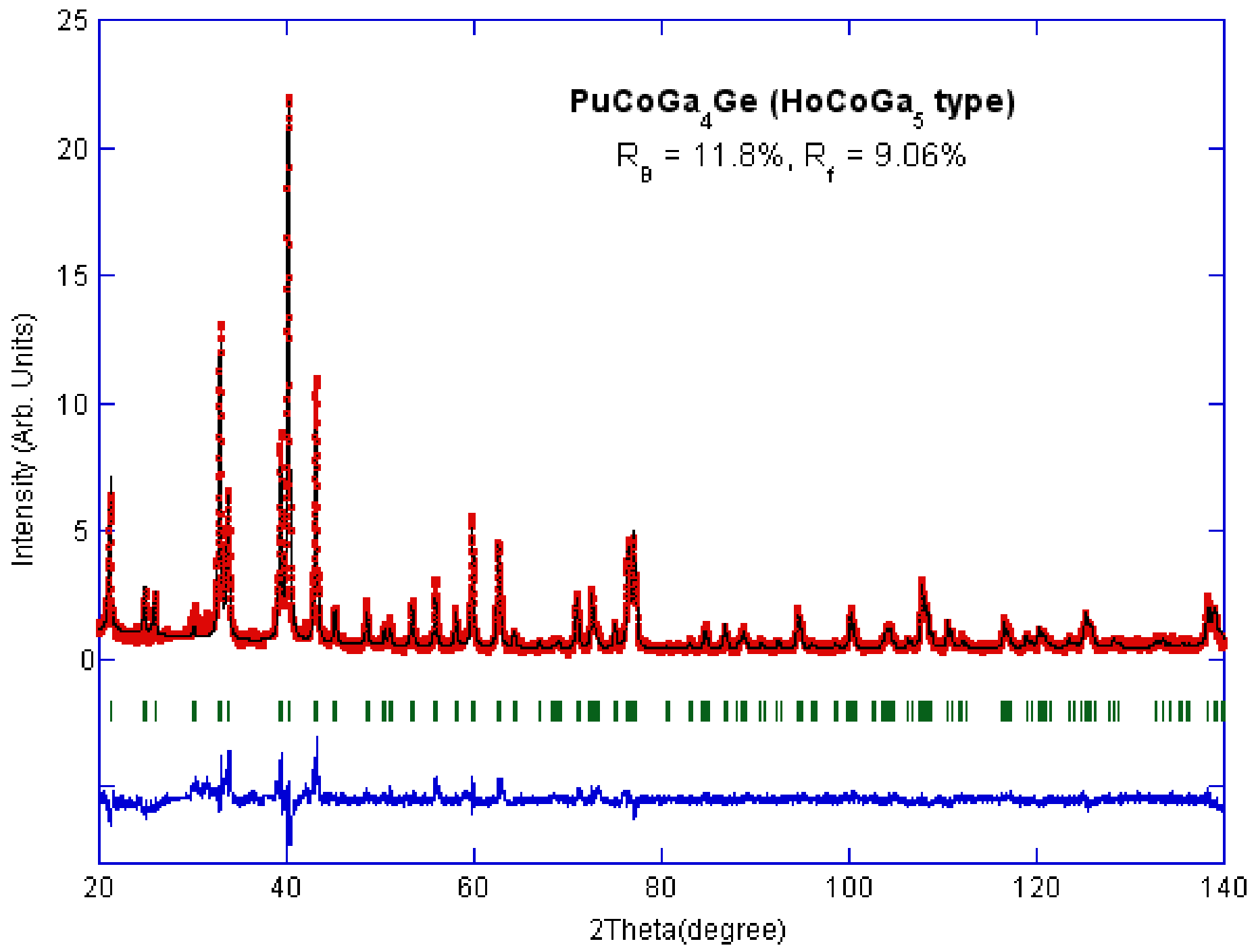

Figure 2

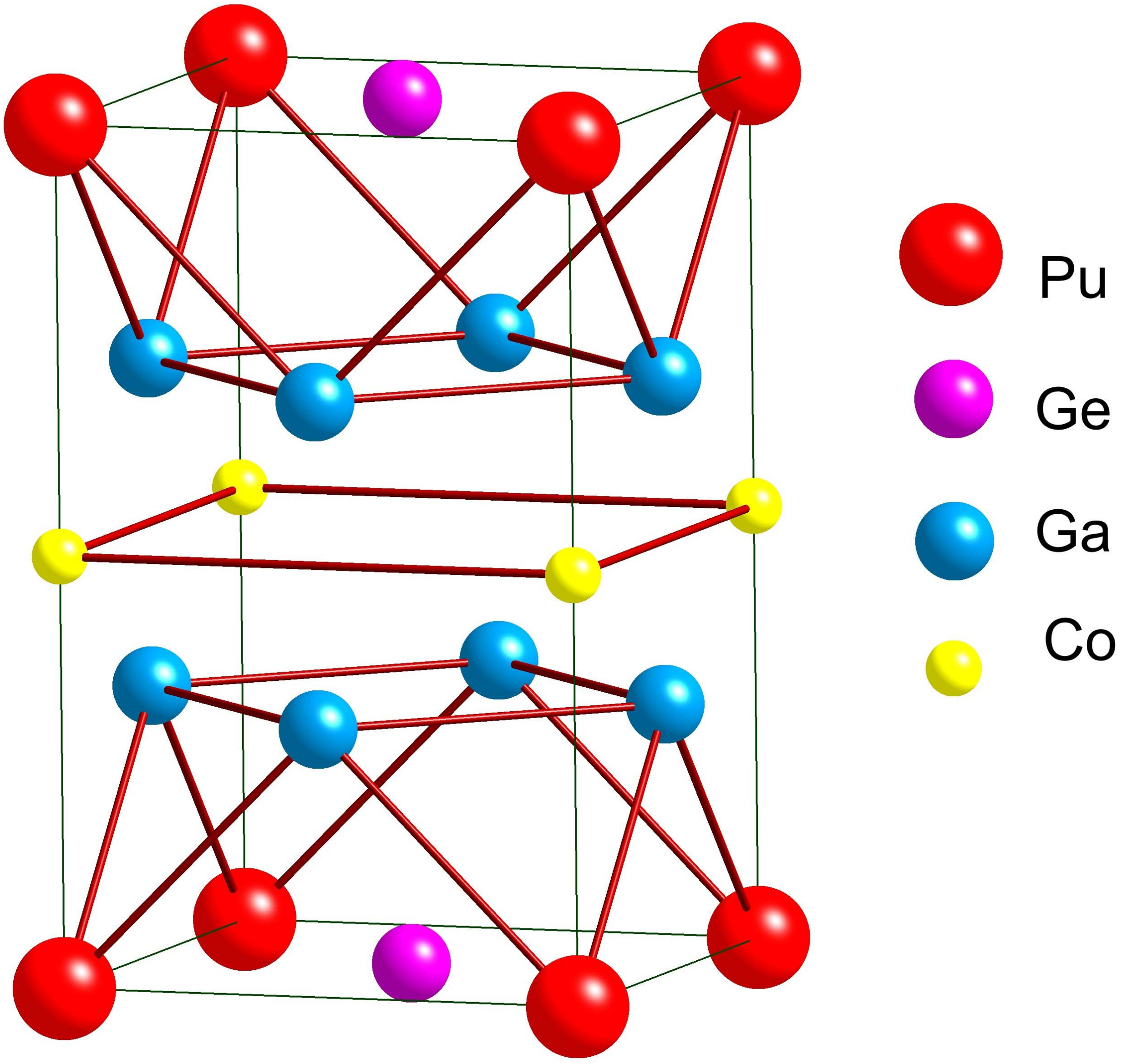

Figure 3

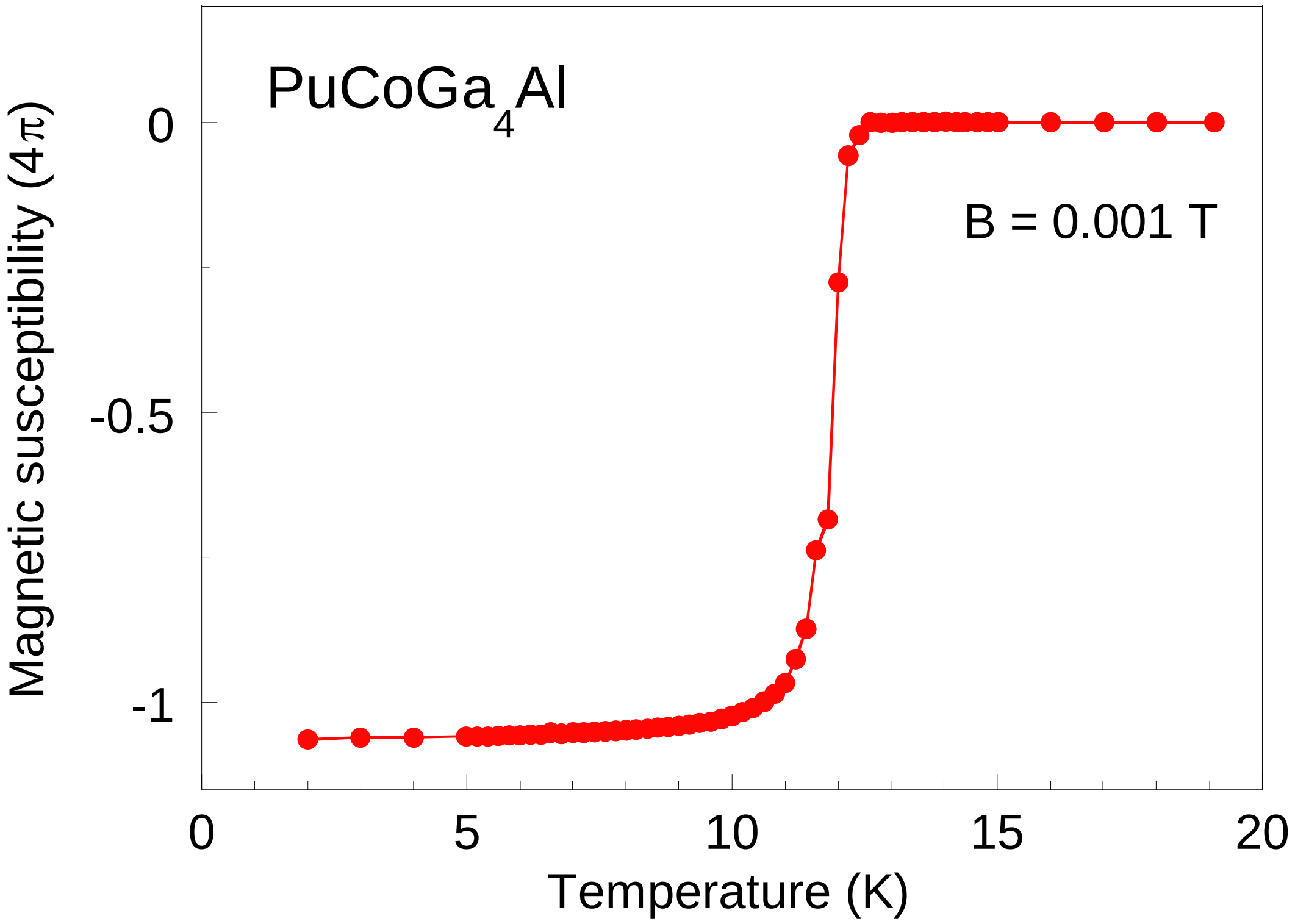

Figure 4

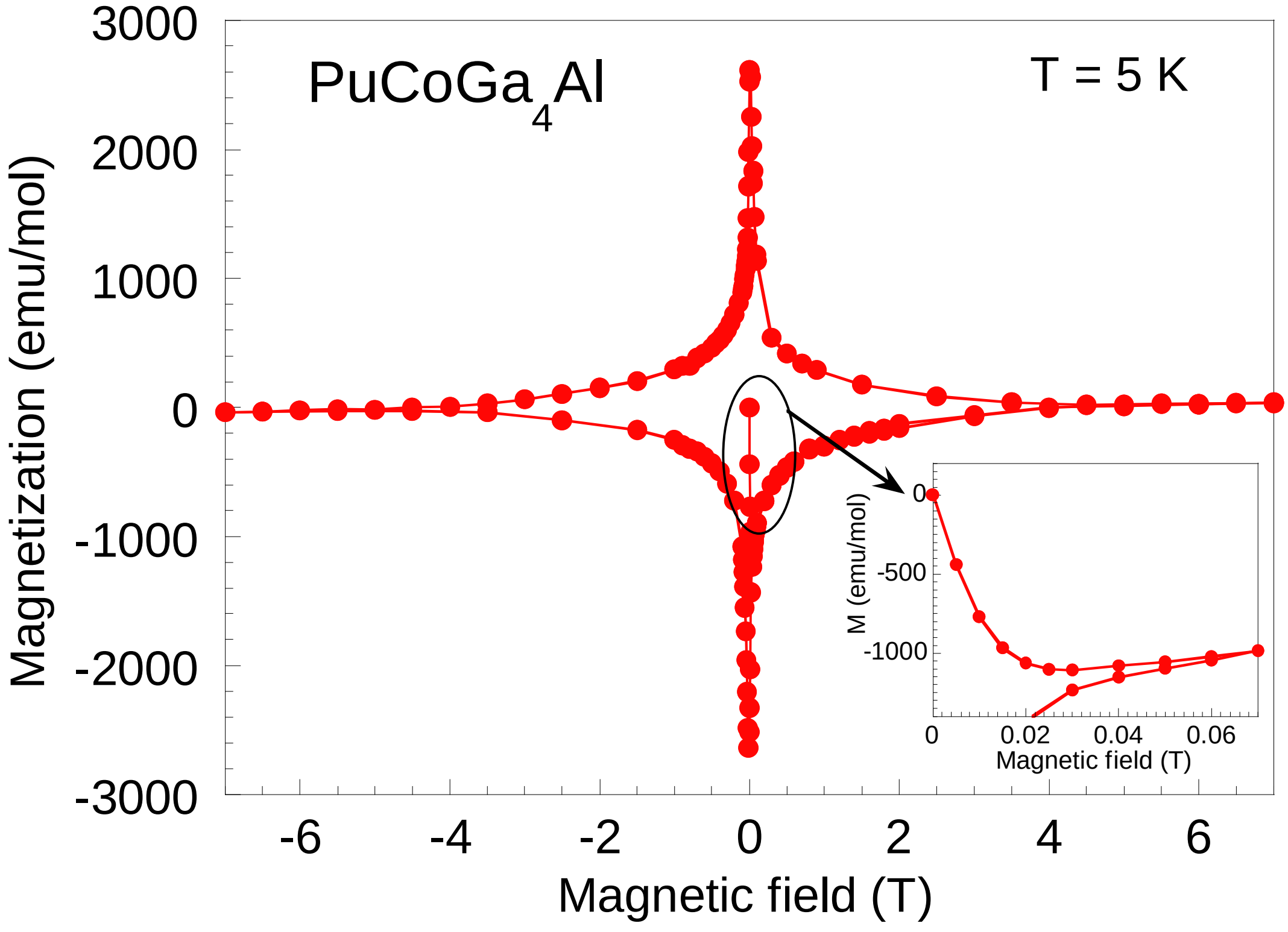

Figure 5

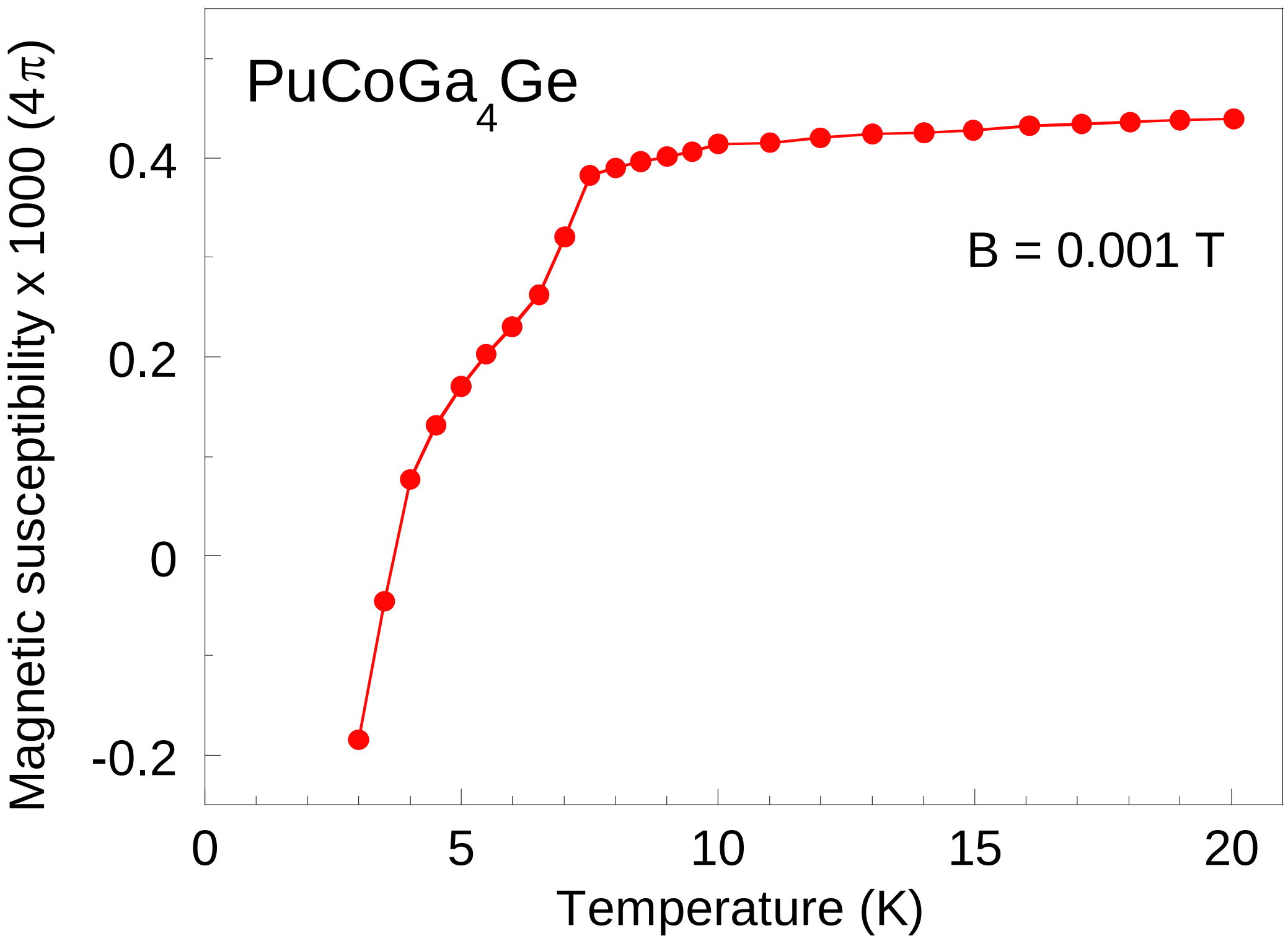

Figure 6

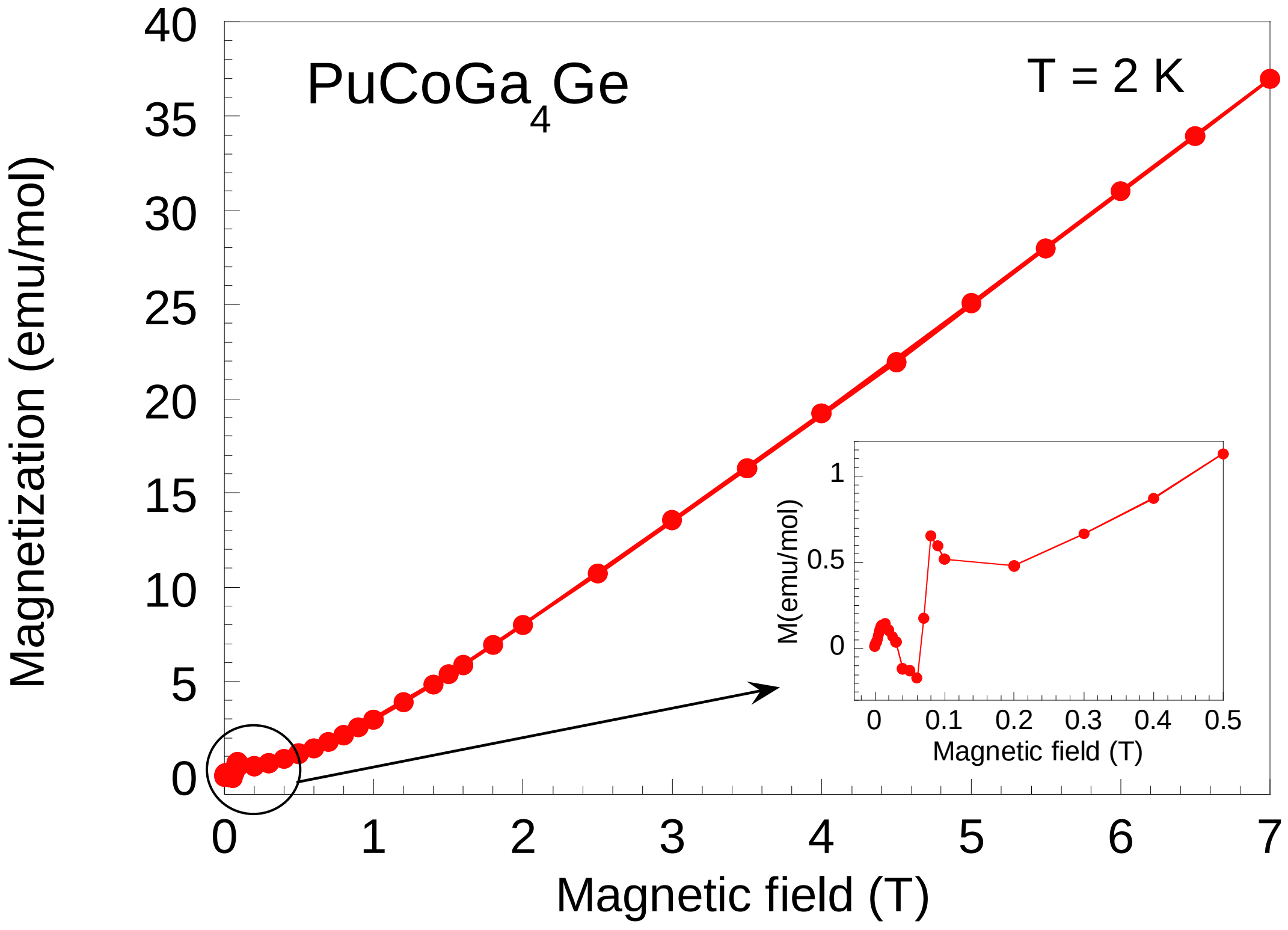

Figure 7

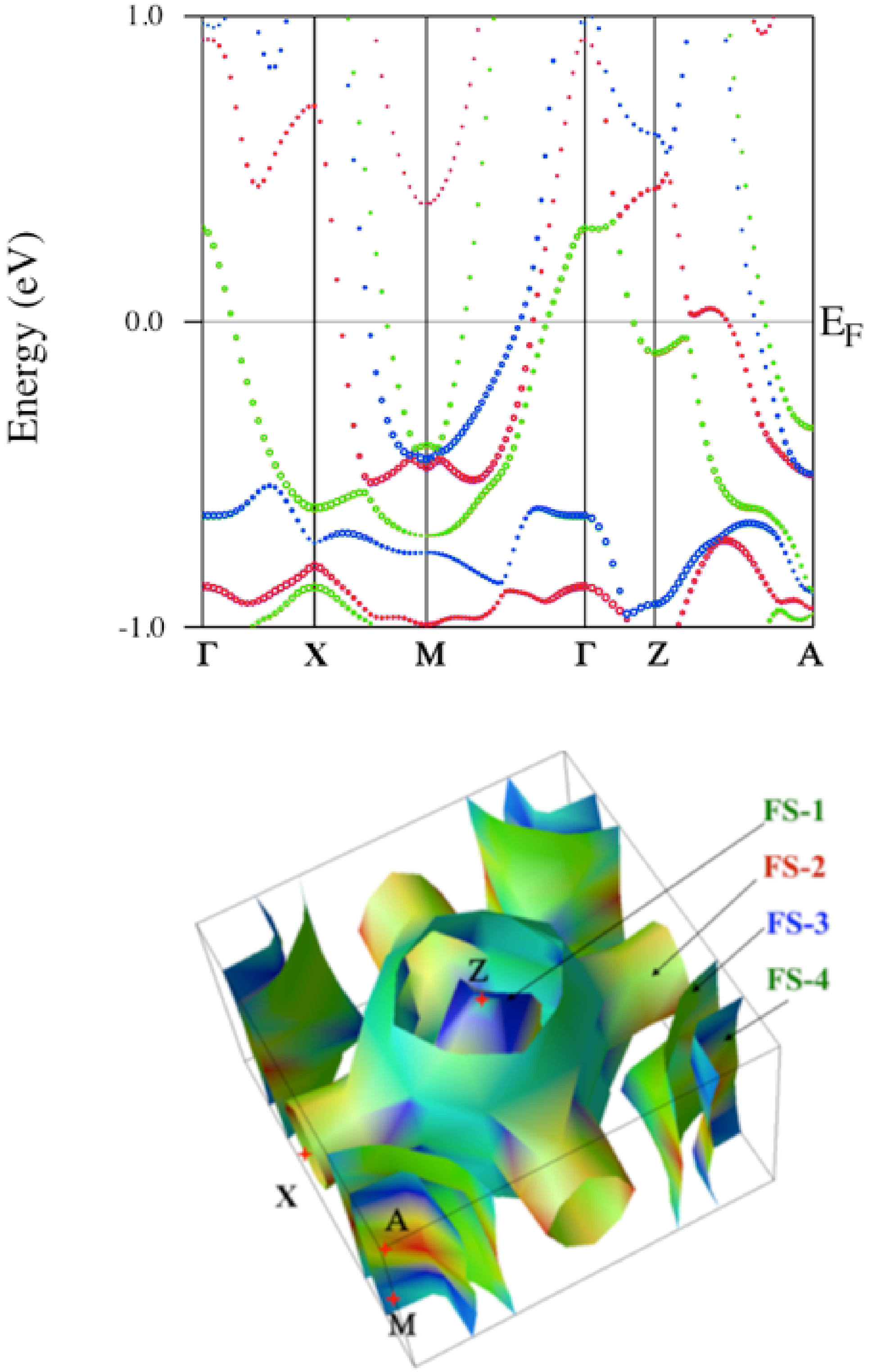

Figure 8

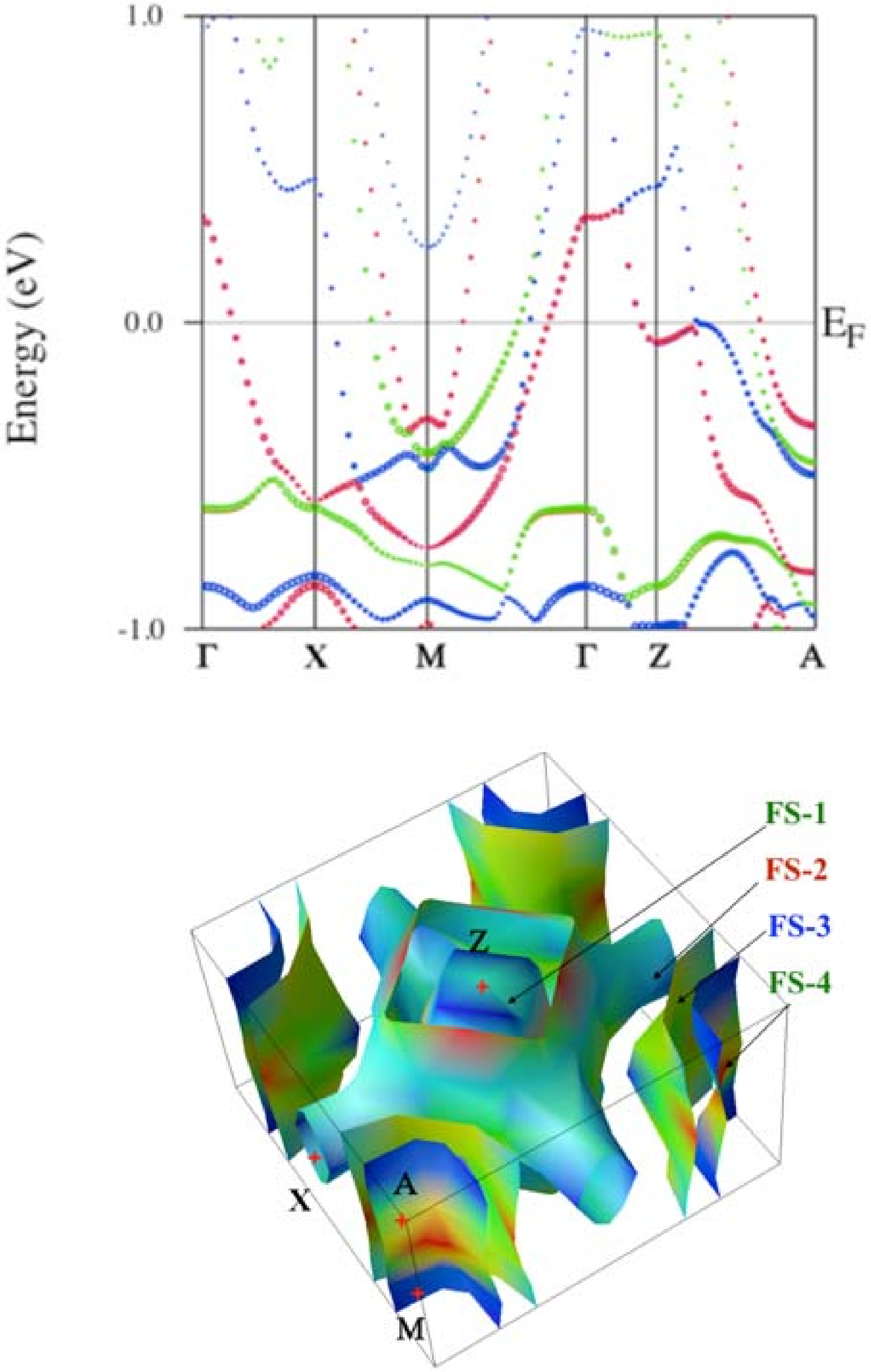

Figure 9

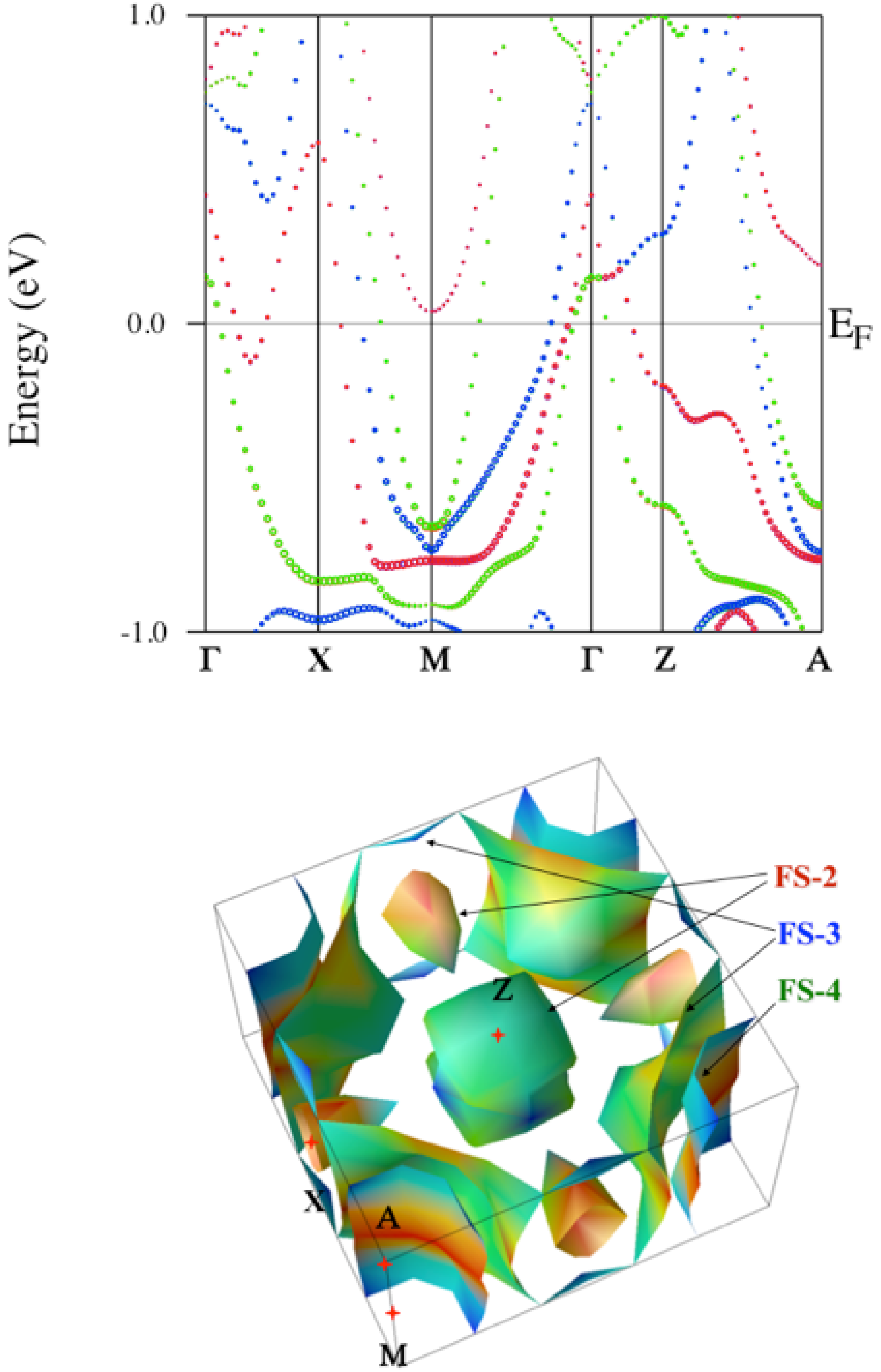

Figure 10

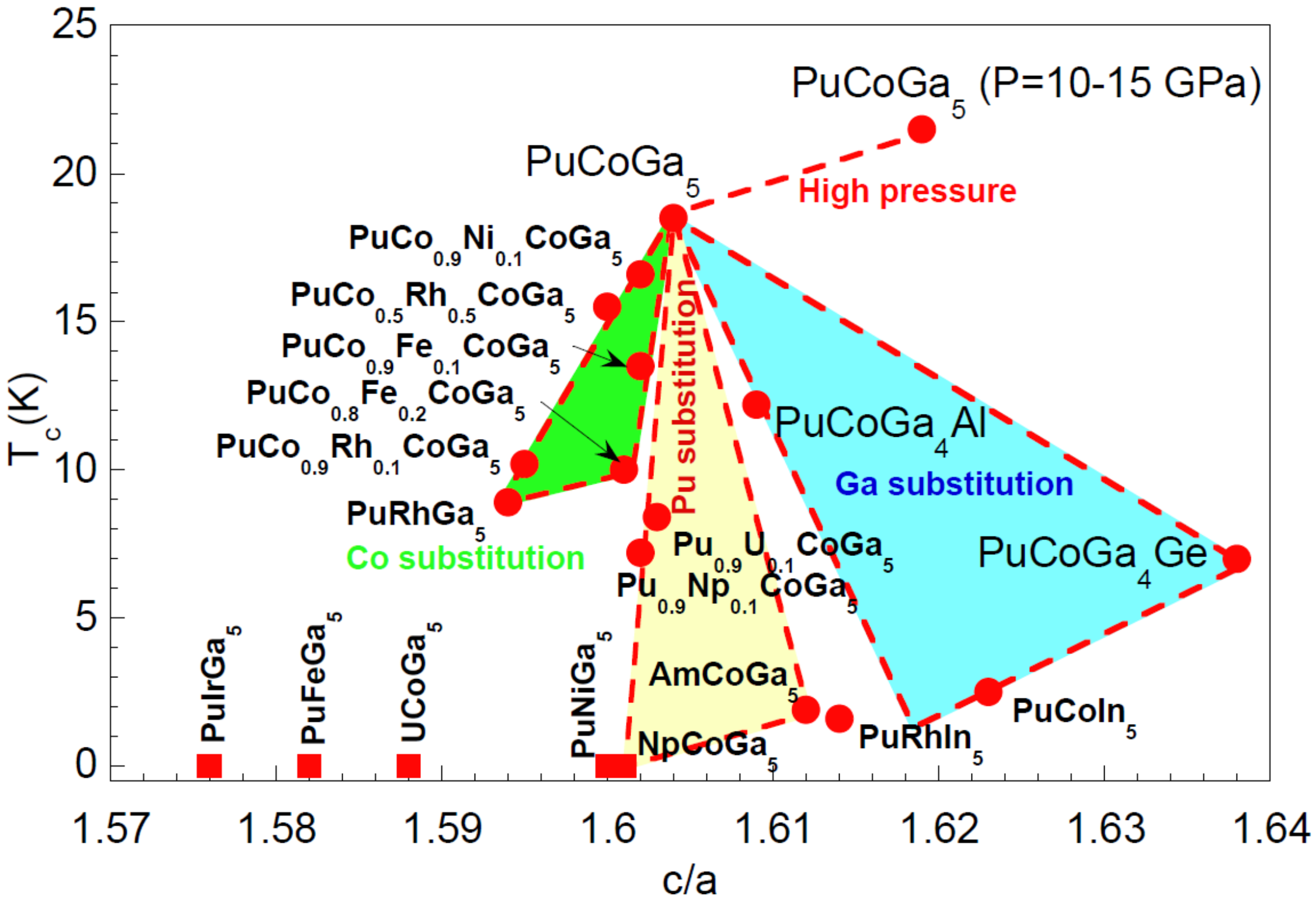

Figure 11